\documentclass[showpacs,amsmath,amssymb,aps,showkeys,floatfix,prd,a4paper]{revtex4}

\usepackage[dvips]{graphicx}

\usepackage{dcolumn}

\usepackage{bm}

\usepackage{epsfig}

\usepackage{amsfonts}

\usepackage{amssymb,amscd}

\usepackage{subfigure}

\usepackage{xcolor}

\newcommand{\rx}{\mbox{\boldmath $x$}}

\newcommand{\ry}{\mbox{\boldmath $y$}}

\newcommand{\rd}{\mbox{\boldmath $\Delta$}}

\def\lsim{\raise0.3ex\hbox{$<$\kern-0.75em\raise-1.1ex\hbox{$\sim$}}}

\def\gsim{\raise0.3ex\hbox{$>$\kern-0.75em\raise-1.1ex\hbox{$\sim$}}}

\newcommand{\be}{\begin{equation}}

\newcommand{\ee}{\end{equation}}

\def\beq{\begin{equation}}

\def\eeq{\end{equation}}

\def\beqa{\begin{eqnarray}}

\def\eeqa{\end{eqnarray}}

\newcommand{\ba}{\begin{eqnarray}}

\newcommand{\rr}{\mbox{\boldmath $r$}}

\newcommand{\rb}{\mbox{\boldmath $b$}}

\def\gappeq{\mathrel{\rlap {\raise.5ex\hbox{$>$}}

{\lower.5ex\hbox{$\sim$}}}}

\def\lappeq{\mathrel{\rlap{\raise.5ex\hbox{$<$}}

{\lower.5ex\hbox{$\sim$}}}}

\def\Toprel#1\over#2{\mathrel{\mathop{#2}\limits^{#1}}}

\begin{document}
\begin{flushright}
LU TP 15-55\\
December 2015
\vskip1cm
\end{flushright}

\title{Exclusive processes  with a leading neutron in $ep$ collisions}

\author{V.P. Gon\c{c}alves$^{1,2}$,  F.S. Navarra$^3$ and D. Spiering$^{3}$}
\affiliation{$^1$ Department of Astronomy and Theoretical Physics, Lund University, SE-223 62 Lund, Sweden \\ 
$^{2}$High and Medium Energy Group, Instituto de F\'{\i}sica e Matem\'atica,  Universidade Federal de Pelotas\\
Caixa Postal 354,  96010-900, Pelotas, RS, Brazil.\\
$^3$Instituto de F\'{\i}sica, Universidade de S\~{a}o Paulo,
C.P. 66318,  05315-970 S\~{a}o Paulo, SP, Brazil.
}

\begin{abstract}
In this paper we extend the color dipole formalism to the study of exclusive processes associated with a leading neutron  in $ep$ collisions 
at high energies. The exclusive $\rho$,  $\phi$ and $J/\Psi$ production, as well as the Deeply Virtual Compton Scattering, are analysed assuming 
a diffractive interaction between the color dipole and the pion emitted by the incident proton. We compare our predictions with the HERA data on  
$\rho$ production and estimate the magnitude of the absorption corrections. We show that the color dipole formalism is able to describe the current 
data. Finally, we present our estimate for the exclusive cross sections which can be studied at HERA and in future electron-proton colliders. 
\end{abstract}



\maketitle

\vspace{1cm}

\section{Introduction}

The study of electron - proton ($ep$) collisions at HERA has improved our understanding of the structure of the proton as well as about the dynamics 
of the strong interactions  at high energies (For a review see e.g. Ref. \cite{paul}). In particular, the study of  diffractive processes has been one 
of the most successful areas at HERA, with  vector meson production and  Deeply Virtual Compton Scattering (DVCS) in exclusive processes 
($\gamma^*p \rightarrow E p$, with  $E = \rho, \phi, J/\Psi, \gamma$) being  important probes of the transition between the soft and hard regimes of  
QCD. These processes have been the subject of intensive theoretical and experimental investigations, with one of the main motivations for these studies 
being the possibility to probe  the QCD dynamics at high energies,  driven by the gluon content of the proton which is strongly subject to non-linear 
effects (parton saturation) \cite{cgc}. An important lesson from the analysis of the HERA data at small values of the Bjorken - $x$ variable is that 
the inclusive and diffractive processes can be satisfactorily  described  using a unified framework -- the color dipole formalism. This approach was 
proposed many years ago in Ref. \cite{nik} and considers that the high energy photon can be described by a color quark - antiquark dipole and that the 
interaction of the dipole with the target  can be described by the color dipole cross section $\sigma_{dt}(x,\rr)$, with the transverse size of the 
dipole $\rr$ frozen during the interaction process.
In this approach all information about the target and strong interaction physics is encoded in 
$\sigma_{dt}(x,\rr)$, which is determined by the imaginary part of the forward amplitude of the scattering between a small dipole
(a colorless quark-antiquark pair) and a dense hadron target, denoted by ${\cal{N}}(x,\rr,\rb)$, 
where the dipole has transverse size given by the vector $\rr=\rx-\ry$, with $\rx$ and $\ry$ being the transverse vectors of the quark
and antiquark, respectively, and $\rb=(\rx+\ry)/2$ is the impact parameter. 
In the Color Glass Condensate (CGC)  formalism \cite{CGC2,BAL},  ${\cal{N}}$ contains all the  
information about  non-linear and quantum effects in the hadron wave function. It can be obtained by solving an appropriate evolution equation in the 
rapidity $Y\equiv \ln (1/x)$, which in its  simplest form is the Balitsky-Kovchegov (BK) equation \cite{BAL,kov}.
Alternatively,  the scattering amplitude can be obtained using phenomenological models based on saturation physics constructed taking into account the 
analytical solutions of the BK equation which are known in the low and high density regimes. As demonstrated in \cite{amir}, the combination between the 
color dipole formalism and saturation physics are quite successful to describe the recent and very precise HERA data on the reduced inclusive cross section 
as well as the data on the exclusive processes in a large range of photon - proton  center - of - mass energies $W$, photon virtualities $Q^2$ and $x$ values.

HERA has also provided high precision experimental data on semi - inclusive  $e + p \rightarrow e + n + X$ processes,  where the incident proton is converted 
into a neutron via a charge exchange \cite{lpdata2}. Very recently  the first measurements of  exclusive $\rho$ photoproduction associated with leading neutrons  
($\gamma p \rightarrow \rho^0 \pi^+ n$) were presented \cite{rhoLN_HERA}.  The description of these leading neutron processes is still a theoretical challenge.
In particular, the  $x_L$ (Feynman momentum) distribution of leading neutrons remains  without a conclusive theoretical description 
\cite{holt,bisha,kope,kuma,niko99,models,kkmr,khoze,speth,pirner}. 
In Ref. \cite{nosLN} we  extended the color dipole formalism to leading neutron processes and demonstrated that the experimental data on the semi - inclusive 
reactions can be well described by this approach. Our goal in this paper is to further extend our previous analysis to exclusive processes and try to show that 
the color dipole formalism may also provide a unified description of leading neutron processes. Using the same assumptions made in Ref. \cite{nosLN},  we 
compare our predictions with the HERA data on $\rho$ exclusive photoproduction and estimate the contribution of the absorption corrections to exclusive 
leading neutron processes. Taking into account these corrections we present  predictions for the exclusive $\phi$, $J/\Psi$ and $\gamma$ production associated
with a leading neutron in $ep$ collisions at the energies of HERA and  future electron - proton colliders.

This paper is organized as follows. In the next Section we present a brief review of  leading neutron production in $ep$ collisions and   we discuss 
the treatment of  exclusive processes with the color dipole formalism. In Section \ref{results} we analyse the dependence of our predictions 
on the models of the vector meson wave function, on  the pion flux  and on  the dipole scattering amplitude. A comparison with the recent HERA data on  
exclusive $\rho$ photoproduction is performed and predictions for  the exclusive $\phi$, $J/\Psi$ and $\gamma$ production associated to a leading neutron in 
$ep$ collisions for the energies of HERA and of the future electron - proton colliders are presented. Finally, in Section  \ref{conc} we summarize our main 
conclusions.

\begin{figure}
\begin{tabular}{cc}
{\psfig{figure=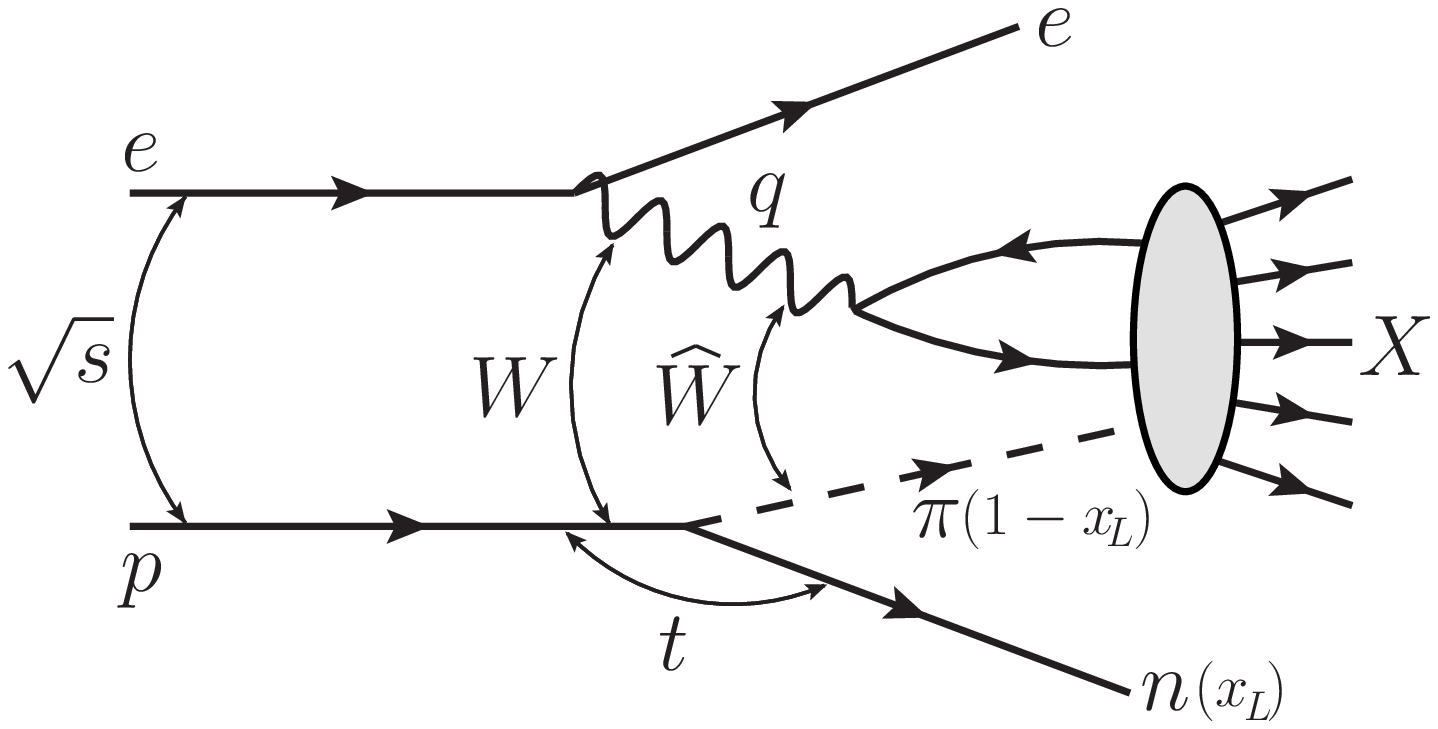,width=7cm}} & {\psfig{figure=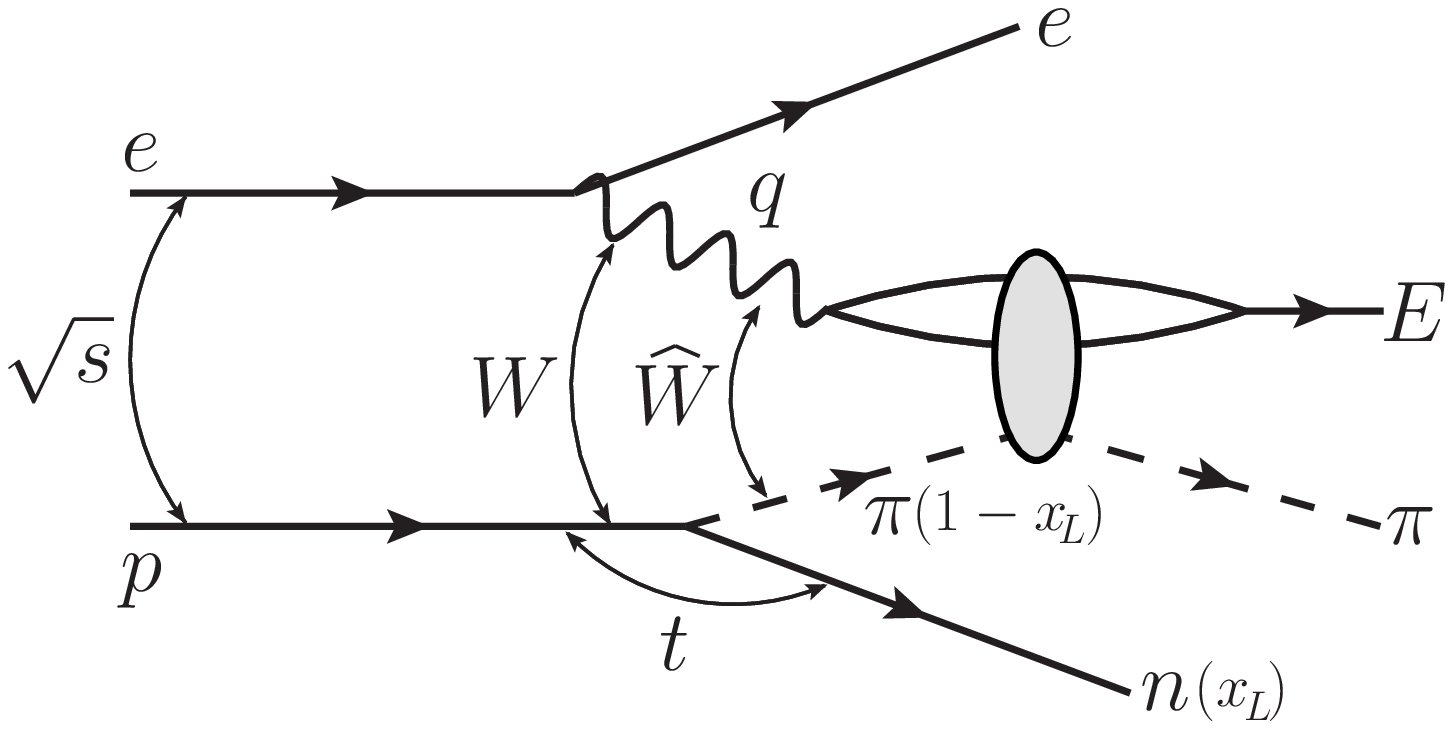,width=7cm}}  
\end{tabular}
\caption{Semi - inclusive (left panel) and exclusive (right panel)  $ep$ processes associated with a leading neutron $n$ production in  the color dipole formalism.}
\label{fig1}
\end{figure}

\section{Exclusive processes associated  with  leading neutron production in the color dipole formalism}
\label{formalismo}

At high energies, the differential cross section for a given process (semi - inclusive or exclusive)  associated with a leading neutron production can 
be expressed as follows:
\beq
\frac{d^2 \sigma(W,Q^2,x_L,t)}{d x_L d t} = f_{\pi/p} (x_L,t) \sigma_{\gamma^* \pi}(\hat{W}^2,Q^2)
\label{crossgen}
\eeq
where $Q^2$ is the virtuality of the exchanged photon, ${W}$ is the center-of-mass energy of the 
virtual photon-proton system, $x_L$ is the proton momentum fraction carried by the 
neutron and $t$ is the square of the four-momentum of the exchanged pion. 
Moreover,  $f_{\pi/p}$ is the flux of virtual pions emitted by the proton  and  $\sigma_{\gamma^* \pi}(\hat{W}^2,Q^2)$  
is the cross section of the interaction between the  virtual-photon and the virtual-pion  at center-of-mass energy $\hat{W}$, which is given by  
$\hat{W}^2 = (1-x_L) \, W^2$. 
The  pion flux $f_{\pi/p} (x_L,t)$ (also called sometimes pion splitting function) is the virtual pion momentum distribution
in a physical nucleon (the bare nucleon plus the ``pion cloud''). In general, it is parametrized as follows 
\cite{holt,bisha,kope,kuma,niko99,models,kkmr,khoze,speth,pirner} 
\beq
f_{\pi/p} (x_L,t)  = \frac{1}{4 \pi} \frac{2 g_{p \pi p}^2}{4  \pi} \frac{-t}{(t-m_{\pi}^2)^2} (1-x_L)^{1-2 \alpha(t)}  
[F(x_L,t)]^2
\label{genflux}
\eeq 
where $g_{p \pi p}^2/(4 \pi) = 14.4$ is the $ \pi^ 0 p p $ coupling constant, $m_{\pi}$ is the pion mass and $\alpha(t)$  is the Regge trajectory of the 
pion. The form factor $F(x_L,t)$  accounts for the finite size of the nucleon and of the pion and is model dependent. As in Ref. \cite{nosLN}, we will consider 
the following parametrizations for the form factor:
\beq
F_1(x_L,t) = \exp \left[ R^2 \frac{(t-m_{\pi}^2)}{(1-x_L)} \right] \,\,\,\, , \,\,\,\, \alpha(t) = 0
\label{form1}
\eeq
from Ref. \cite{holt}, where $R = 0.6$ GeV$^{-1}$. 
\beq
F_2(x_L,t) = 1 \,\,\,\, , \,\,\,\, \alpha(t) = \alpha(t)_{\pi}
\label{form2}
\eeq
from  Ref. \cite{bisha}, where $\alpha_{\pi}(t) \simeq t$ (with $t$ in GeV$^2$) is the Regge trajectory of the pion.
\beq
F_3(x_L,t) =  \exp \left[ b (t-m_{\pi}^2) \right] \,\,\,\, , \,\,\,\, \alpha(t) = \alpha(t)_{\pi} 
\label{form3}
\eeq
from Ref. \cite{kope}, where $\alpha_{\pi}(t) \simeq t$ (with $t$ in GeV$^2$)  and $b = 0.3$ GeV$^{-2}$. 
\beq
F_4(x_L,t) =  \frac{\Lambda_m^2-m_{\pi}^2}{\Lambda_m^2-t}      \,\,\,\, , \,\,\,\, \alpha(t) = 0
\label{form4}
\eeq
from Ref. \cite{kuma}, where $\Lambda_m = 0.74$ GeV. 
\beq
F_5(x_L,t) =  \left[\frac{\Lambda_d^2-m_{\pi}^2}{\Lambda_d^2-t}\right]^2      \,\,\,\, , \,\,\,\, \alpha(t) = 0
\label{form5}
\eeq
also from Ref. \cite{kuma}, where $\Lambda_d = 1.2$ GeV. In what follows we will denote the corresponding pion flux associated to these different 
form factors by 
$f_1$, $f_2$,  ... $f_5$, respectively. Moreover, it is important to emphasize that in the case of the more familiar exponential (\ref{form1}), 
monopole (\ref{form4})  and dipole (\ref{form5}) 
forms factors, the cut-off parameters have been determined by fitting low energy data on nucleon and nuclear reactions and also data on deep inelastic 
scattering and structure functions \cite{clouds}.

In Ref. \cite{nosLN}, we  described the semi - inclusive leading neutron processes in the color dipole formalism. The basic idea is that at high energies, 
this process can be seen as a sequence of three factorizable subprocesses [See Fig. \ref{fig1} (left panel)]:  i) the photon fluctuates into a 
quark-antiquark pair (the color dipole), ii) the color dipole interacts with the pion, present in the wave function of the incident 
proton, and iii) the leading neutron is formed. Consequently, the photon - pion cross section can be  factorized  in terms of the photon wave functions 
$\Psi$, which describes the photon splitting in a $q\bar{q}$ pair, and the dipole-pion cross section  $\sigma_{d\pi}$. 
In the eikonal approximation the  
dipole-proton cross section $\sigma_{d\pi}$  is given by:
\begin{equation} 
\sigma_{d\pi} (\hat{x}, \rr) = 2 \int d^2 \rb \,  {\cal N}^\pi (\hat{x}, \rr, \rb)\,\,,
\label{sdip}
\end{equation}
where
\beq
\hat{x} = \frac{Q^2 + m^2_f}{\hat{W}^2 + Q^2} = \frac{Q^2 + m^2_f}{(1-x_L)W^2 + Q^2}
\label{xhat}
\eeq
is the scaled Bjorken variable and  $\mathcal{N}^\pi(x,\rr,\rb)$ is  the imaginary part of the forward amplitude of the scattering between a small dipole
(a colorless quark-antiquark pair) and a pion, at a given
rapidity interval $Y=\ln(1/\hat{x})$. In Ref. \cite{nosLN} we proposed to relate 
$\mathcal{N}^\pi$ with the dipole-proton scattering amplitude  $\mathcal{N}^p$, usually probed 
in the typical inclusive and exclusive processes at HERA, assuming that 
\begin{equation}
{\cal N}^\pi (\hat{x}, \rr, \rb) = R_q \cdot {\cal N}^p (\hat{x}, \rr, \rb) 
\label{doister}
\end{equation}
with $R_q$ being a constant. In the additive quark model it is expected that $R_q = 2/3$, which is the ratio between the number of valence quarks in the 
target hadrons. 
{This model was first applied to soft hadronic reactions \cite{lefra} and, in particular,  it predited the following relation between the pion-proton and 
proton-proton total cross sections: $\sigma_{\pi p} = 2/3 \,  \sigma_{pp}$. This relation 
is observed experimentally. It refers to  total cross sections and the only kinematical variable is the c.m.s energy $\sqrt{s}$. In the low energy 
domain, where it was verified, the dependence on $\sqrt{s}$ was very weak. As it was discussed in Ref. \cite{kpps},  in hard hadronic reactions, where a high 
energy scale is present, Eq. (\ref{doister}) may still be valid, although  deviations from $2/3$ are likely to be seen. The idea is that dipoles with 
$Q^2 \ge 10$ GeV$^2$ can resolve the quarks in the target and interact with each of them independently. The cross section is then proportional to the number of 
quarks in the target. At increasing $Q^2$ and/or collision energies quantum fluctuations become more important, increasing the effective number of quarks. 
According to Ref. \cite{kpps}, this growth is stronger in the proton than in the pion and hence $2/3 \rightarrow 1/2$ (or even $1/3$). Here we consider a range 
of $R_q$, going from $1/3$ up to $2/3$. As it will be seen, our results imply that if $R_q =2/3$ the absorption factor $K$ tends to be too small. In view of the 
existing calculations of $K$, we would  conclude that  $R_q =2/3$ is probably too large. }
Since the effective value of this quantity is still  an open question \cite{zoller,kkmr,kpps,khoze}, we have considered in  \cite{nosLN}  
that $R_q$ could be in the range $1/3 \le R_q \le 2/3$. With this basic assumption we have estimated the dependence of the predictions on the description 
of the QCD dynamics at high energies as well as the contribution of  gluon saturation effects to  leading neutron production. Moreover, with the 
parameters  constrained by other phenomenological information, we were able to reproduce the basic features of  the  
H1 data on  leading neutron spectra \cite{lpdata2}.

As mentioned in Ref. \cite{nosLN}, one source of uncertainty in the study of inclusive leading neutron production (in Fig. \ref{fig1} on the left) is the 
fact that there are several processes which lead to the same final state. Apart from one pion emission we may have, for example, $\rho$ emission. Even with 
pion emission we may have $\Delta$  production with the subsequent decay $\Delta \rightarrow n + \pi$. The strength of these contributions is highly model  
dependent and their existence prevents us from extracting more precise information on the photon-pion cross section or on the pion flux.  
In contrast, in  $\rho$ exclusive production with  a leading neutron 
none of these processes contributes to the exclusive reaction shown in the right panel of Fig. \ref{fig1}.  This feature makes the leading neutron spectrum 
measured in exclusive processes a 
better testing ground for both the determination 
of the photon-pion cross section and of the pion flux.

In what follows we will assume that the factorization given by Eq. (\ref{crossgen}) also is valid  and that the photon - pion cross section for the production 
of an exclusive final state $E$, such as a vector 
meson ($E = V$) or a real photon in DVCS ($E = \gamma$), in the  $\gamma^* \pi \rightarrow E \pi$ process is given in the color dipole formalism by:
\begin{eqnarray}
\sigma (\gamma^* \pi \rightarrow E \pi) = \sum_{i=L,T} \int_{-\infty}^0 \frac{d\sigma_i}{d\hat{t}}\, d\hat{t}  
= \frac{1}{16\pi} \sum_{i=L,T} \int_{-\infty}^0 |{\cal{A}}_i^{\gamma^* \pi \rightarrow E \pi }(x,\Delta)|^2 \, d\hat{t}\,\,,
\label{sctotal_intt}
\end{eqnarray}
with the scattering amplitude being given by 
 \begin{eqnarray}
 {\cal A}_{T,L}^{\gamma^* \pi \rightarrow E \pi}(\hat{x},\Delta)  =  i
\int dz \, d^2\rr \, d^2\rb  e^{-i[\rb-(1-z)\rr].\rd} 
 \,\, (\Psi^{E*}\Psi)_{T,L}  \,\,2 {\cal{N}}_\pi(\hat{x},\rr,\rb)
\label{sigmatot2}
\end{eqnarray}
where $(\Psi^{E*}\Psi)_{T,L}$ denotes the overlap of the photon and exclusive final state wave functions. The variable  $z$ $(1-z)$ is the
longitudinal momentum fractions of the quark (antiquark) and  $\Delta$ denotes the transverse 
momentum lost by the outgoing pion ($\hat{t} = - \Delta^2$). 
The variable $\rb$ is the transverse distance from the center of the target to the center of mass of the $q \bar{q}$  dipole and the factor  in the 
exponential  arises when one takes into account 
non-forward corrections to the wave functions \cite{non}.
In what follows we will assume that  the vector meson is predominantly a quark-antiquark state 
and that the spin and polarization structure is the same as in the  photon \cite{dgkp,nnpz,sandapen,KT} (for other approaches see, for example, 
Ref. \cite{pacheco}). As a consequence, the overlap between the photon and the vector meson wave function, for the transversely and longitudinally 
polarized  cases, is given by (For details see Ref. \cite{kmw})
\begin{eqnarray}
  (\Psi^*_V\Psi)_T &=& \frac{\hat e_fe}{4\pi}\frac{N_c}{\pi z(1-z)}
    \left\{m_f^2K_0(\epsilon r)\phi_T(r,z)-\left[z^2+(1-z)^2\right]\epsilon K_1(\epsilon r)\partial_r\phi_T(r,z)\right\}, \\
  (\Psi^*_V\Psi)_L &=& \frac{\hat e_fe}{4\pi}\frac{N_c}{\pi}2Qz(1-z)K_0(\epsilon r) 
    \left[M_V\phi_L(r,z) + \delta \frac{m_f^2-\nabla_r^2}{M_Vz(1-z)} \phi_L(r,z) \right],
\end{eqnarray}
where $ \hat{e}_f $ is the effective charge of the vector meson, $m_f$ is the quark mass, $N_c = 3$, $\epsilon^2 = z(1-z)Q^2 + m_f^2$   
and $\phi_i(r,z)$ define the scalar parts of the  vector meson wave functions. We will consider the Boosted Gaussian and 
Gauss-LC models for $\phi_T(r,z)$ and $\phi_L(r,z)$, which are largely used in the literature.
In the Boosted Gaussian model the functions $\phi_i(r,z)$ are given by 
\begin{eqnarray}
  \phi_{T,L}(r,z) =  {\cal{C}}_{T,L}\,z(1-z)\exp
  \left[ -\frac{m_f^2 R^2}{8z(1-z)} - \frac{2z(1-z)r^2}{ R^2} + \frac{m_f^2 {R}^2}{2} \right].
\end{eqnarray}
 In contrast, in the Gauss-LC model, they are given by
\begin{eqnarray}
  \phi_T(r,z) &=& N_T\left[z(1-z)\right]^2\exp\left(-r^2/2R_T^2\right),\\
  \phi_L(r,z) &=& N_L z(1-z)\exp\left(-r^2/2R_L^2\right).
\end{eqnarray}
The parameters ${\cal{C}}_i$,  $R$, $N_i$  and $R_i$ are  determined by the normalization condition of the wave function and by the meson decay width. 
In Table \ref{tab:1}  we present the value of these parameters for the vector meson wave functions. It is important to emphasize that predictions 
based on these models for the wave functions have been tested with success in $ep$ and ultra peripheral hadronic collisions (See, e. g. 
Refs. \cite{amir,anelise,vicper}).
In the DVCS case, as one has a 
real photon at the final state, only the transversely polarized overlap function contributes 
to the cross section.  Summed over the quark helicities, for a given quark flavour $f$ it is 
given by \cite{kmw},
\begin{widetext}
\begin{eqnarray}
  (\Psi_{\gamma}^*\Psi)_{T}^f & = & \frac{N_c\,\alpha_{\mathrm{em}}
e_f^2}{2\pi^2}\left\{\left[z^2+\bar{z}^2\right]\varepsilon_1 K_1(\varepsilon_1 r) 
\varepsilon_2 K_1(\varepsilon_2 r) 
 +     m_f^2 K_0(\varepsilon_1 r) K_0(\varepsilon_2 r)\right\},
  \label{eq:overlap_dvcs}
\end{eqnarray}
\end{widetext}
where we have defined the quantities $\varepsilon_{1,2}^2 = z\bar{z}\,Q_{1,2}^2+m_f^2$ and 
$\bar{z}=(1-z)$. Accordingly, the photon virtualities are $Q_1^2=Q^2$ (incoming virtual 
photon) and $Q_2^2=0$ (outgoing real photon).

\begin{table}[t]
  \centering
  \begin{tabular}{|ccccccccccc|}
    \hline
    Meson    & $M_V$/GeV & $m_f$/GeV & $\hat{e}_f$ & $N_T$ & ${\cal{C}}_T$ & $R_T^2$/GeV$^{-2}$ & $N_L$ & ${\cal{C}}_L$ & $R_L^2$/GeV$^{-2}$ & $R^2$/GeV$^{-2^{\hspace{-0.1cm}\phantom{|}}}$\\ 
    \hline
    $\rho$   & 0.776     & 0.14      & 1/$\sqrt{2}^{\phantom{|}}$& 4.47  & 0.911      & 21.9               & 1.79  & 0.853      & 10.4              & 12.9                 \\
    $\phi$   & 1.019     & 0.14      & 1/3         & 4.75  & 0.919      & 16.0               & 1.41  & 0.825      & 9.7               & 11.2                 \\
    $J/\psi$ & 3.097     & 1.4       & 2/3         & 1.23  & 0.578      & 6.5                & 0.83  & 0.575      & 3.0               &  2.3                 \\
    \hline
  \end{tabular}
  \caption{Parameters for the Boosted Gaussian and Gauss-LC wave functions for the different vector mesons.}
  \label{tab:1}
\end{table}

Finally, in order to estimate the photon - pion cross section we must specify the dipole - pion 
scattering amplitude ${\cal N}^\pi$. As considered in Ref. \cite{nosLN} for the semi - inclusive processes, 
we will assume the validity of the approximation expressed by Eq. (\ref{doister}), with the dipole proton 
scattering amplitude ${\cal N}^p$ being given by the bCGC model, proposed in 
Ref.  \cite{kmw} and recently updated in 
Ref. \cite{amir}, which is  based on the CGC formalism and takes into account the impact parameter dependence 
of the dipole - proton scattering amplitude.  As demonstrated in Refs. \cite{amir,vicper}, this model is able to describe the vector meson 
production in $ep$ and ultra peripheral hadronic collisions. In the bCGC model the dipole - proton scattering amplitude is given by \cite{kmw} 
\begin{widetext}
\begin{eqnarray}
\mathcal{N}^p(\hat{x},\rr,\rb) =   
\left\{ \begin{array}{ll} 
{\mathcal N}_0\, \left(\frac{ r \, Q_s(b)}{2}\right)^{2\left(\gamma_s + 
\frac{\ln (2/r Q_s(b))}{\kappa \,\lambda \,Y}\right)}  & \mbox{$r Q_s(b) \le 2$} \\
 1 - e^{-A\,\ln^2\,(B \, r  Q_s(b))}   & \mbox{$r Q_s(b)  > 2$} 
\end{array} \right.
\label{eq:bcgc}
\end{eqnarray}
\end{widetext} 
with  $\kappa = \chi''(\gamma_s)/\chi'(\gamma_s)$, where $\chi$ is the 
LO BFKL characteristic function.  The coefficients $A$ and $B$  
are determined uniquely from the condition that $\mathcal{N}^p(\hat{x},\rr,\rb)$, and its derivative 
with respect to $r\,Q_s(b)$, are continuous at $r\,Q_s(b)=2$. 
In this model, the proton saturation scale $Q_s(b)$ depends on the impact parameter:
\begin{equation} 
  Q_s(b)\equiv Q_s(\hat{x},b)=\left(\frac{x_0}{\hat{x}}\right)^{\frac{\lambda}{2}}\;
\left[\exp\left(-\frac{{b}^2}{2B_{\rm CGC}}\right)\right]^{\frac{1}{2\gamma_s}}.
\label{newqs}
\end{equation}
The parameter $B_{\rm CGC}$  was  adjusted to give a good 
description of the $t$-dependence of exclusive $J/\psi$ photoproduction.  
Moreover, the factors $\mathcal{N}_0$ and  $\gamma_s$  were  taken  to be free.  The set of parameters  which will be used here is the 
following: $\gamma_s = 0.6599$, $\kappa = 9.9$, $B_{CGC} = 5.5$ GeV$^{-2}$, $\mathcal{N}_0 = 0.3358$, $x_0 = 0.00105$ and $\lambda = 0.2063$.  
Moreover, in order to estimate the dependence of our predictions on the choice of the model for ${\cal N}^p$, we  also will consider the IIMS
 \cite{iim,soyez} and GBW \cite{GBW} models, as well as the numerical solution of the BK equation obtained in Ref. \cite{bkrunning}. Such models 
were discussed in detail in  Ref. \cite{nosLN}.
For these models, we assume $\mathcal{N}^p(\hat{x},\rr,\rb) = \mathcal{N}^p(\hat{x},\rr) S(\rb)$ and 
$\sigma_{dp} (\hat{x}, \rr) = \sigma_0 \cdot  {\cal N}^p (\hat{x}, \rr)$, with the normalization of the dipole cross section ($\sigma_0$) being 
fitted to data, and that the $\hat{t}$ - dependence of the photon - pion cross section  can be approximated by an exponential ansatz, 
$d\sigma/d\hat{t} = d\sigma/d\hat{t} (\hat{t} =0) \cdot e^{-B |\hat{t}|}$,  with the slope being given by  $B = \sigma_0 / 4 \pi$.  It is important to 
emphasize that the conclusions obtained in \cite{nosLN} are not modified if the bCGC model is used as input in the calculations.

Before discussing our results, a comment is in order. As in the semi - inclusive case, our predictions for the exclusive 
processes associated with a leading neutron are essentially parameter free, depending only on the choices of the models for the pion flux and on the 
dipole scattering amplitude. The main uncertainties are associated with the choice of $R_q$ (in Eq. (\ref{doister})) and  the magnitude of the absorption 
effects which can arise by soft rescatterings. These latter  are difficult to calculate \cite{speth,pirner} but are expected to modify  almost uniformly 
all the $x_L$ spectrum of the leading neutrons. 
As in Ref. \cite{nosLN}, in what follows we will assume that these effects can be mimicked by a factor $K$, which multiplies the right side of Eq. 
(\ref{crossgen}) changing the normalization of the spectra and which should be estimated from the analysis of experimental data.  In spite of the efforts made in
several studies of absorptive corrections  in semi - inclusive processes \cite{speth,pirner,kkmr,khoze,kpps,kpss},   the 
magnitude of these effects in exclusive processes remains an open question.

\begin{figure}[t]
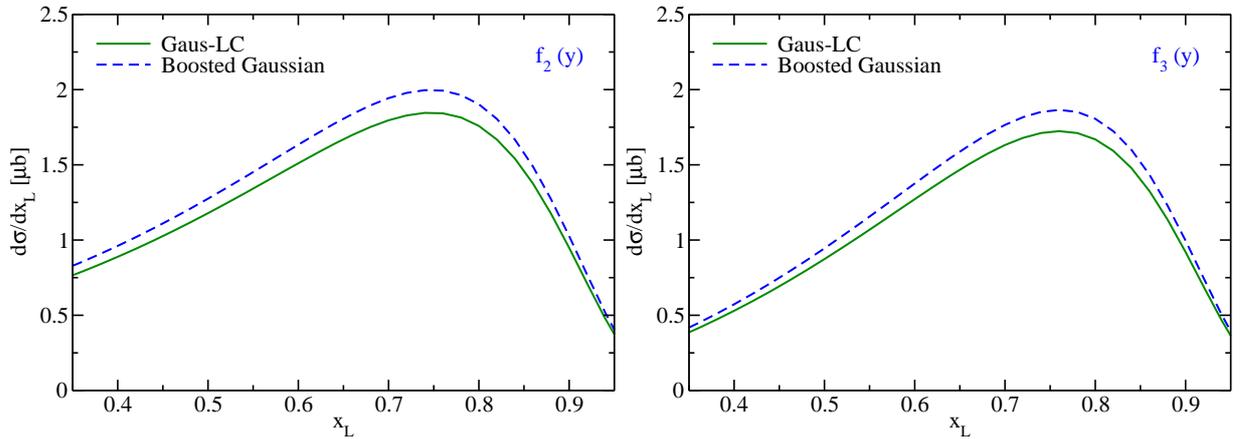

 \centering
 \includegraphics[scale=0.33]{wave_bCGC_f2.eps}
 \includegraphics[scale=0.33]{wave_bCGC_f3.eps}
 \caption{Leading neutron spectra in  exclusive $\rho$ photoproduction considering two different models for the vector meson wave function 
(Boosted Gaussian and Gauss - LC) and two different models for the pion flux ($f_2$ and $f_3$).}
 \label{fig:2}
\end{figure}

\section{Results}

\label{results}

\begin{figure}[t]
 \centering
 \includegraphics[scale=0.4]{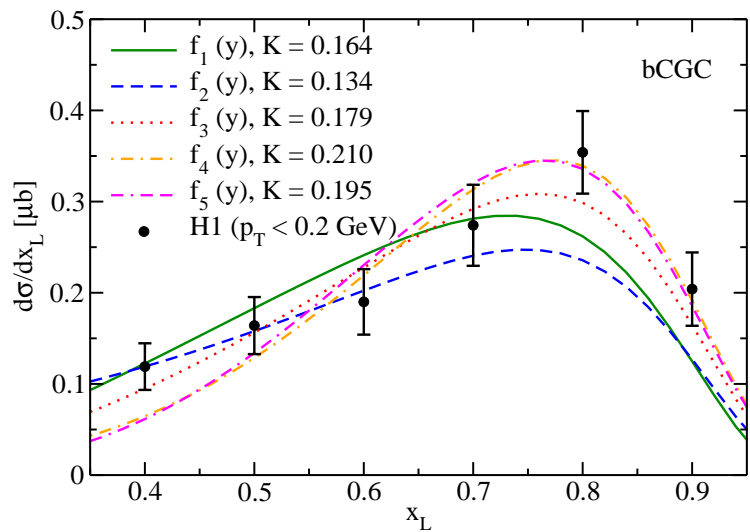}
 \caption{Leading neutron spectra in exclusive $\rho$ photoproduction considering different models of the pion flux. Data from Ref. \cite{rhoLN_HERA}.}
 \label{fig:3}
\end{figure}

\begin{figure}[t]
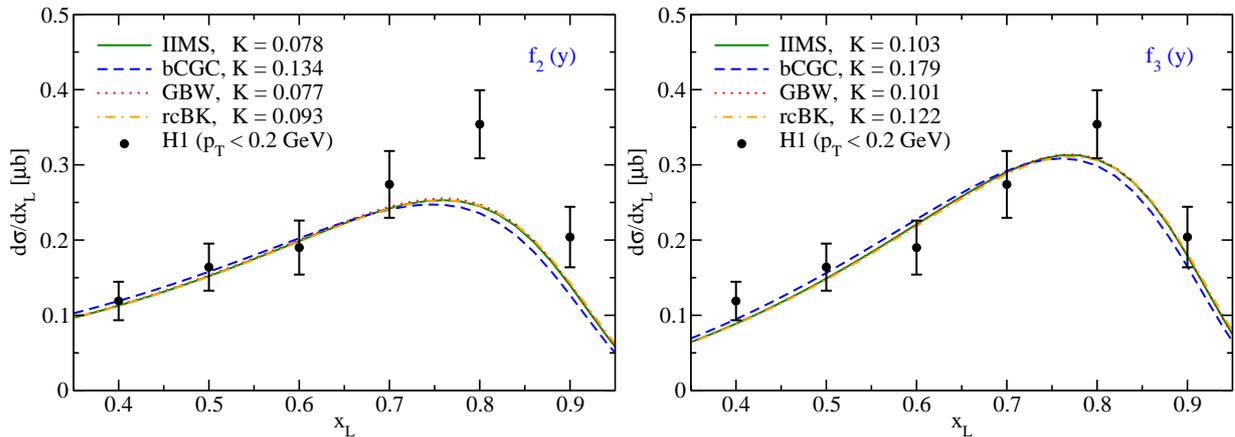

 \centering
 \includegraphics[scale=0.33]{sigma_dip_f2.eps}
 \includegraphics[scale=0.33]{sigma_dip_f3.eps}
 \caption{Leading neutron spectra in exclusive $\rho$ photoproduction considering different models for the dipole scattering amplitude and of  the pion 
flux. Data from Ref. \cite{rhoLN_HERA}.}
 \label{fig:4}
\end{figure}

Let us start our analysis considering the exclusive $\rho$ photoproduction 
associated with leading neutrons as analysed by the H1 Collaboration \cite{rhoLN_HERA}. In what follows we will assume that $W = 60$ GeV, $Q^2 = 0.04$ GeV$^2$ 
and that $p_T < 0.2$ GeV, where $p_T$ is the transverse momentum of the leading neutron. Moreover, we will assume initially that $R_q = 2/3$ and the bCGC 
dipole model. In Fig. \ref{fig:2} we analyse the dependence of our predictions  on the choice of the vector meson wave 
function. We present our results for two different models of the pion flux. We find that the predictions are similar, with the Gauss-LC results being a 
lower bound. This conclusion is also valid for other models of the pion flux and for the $\phi$ and $J/\Psi$ production. Consequently, in what follows we 
will consider only the Gauss-LC model for the vector meson wave functions. Let us now compare our predictions with the experimental data \cite{rhoLN_HERA} 
considering different models for the pion flux. In order to constrain the value of the $K$ factor associated to absorptive corrections,  our strategy will 
be following: for a given  model of the pion flux, $R_q$ and dipole cross section, we will estimate the total cross section. The value of $K$ will be the value 
necessary to make our prediction consistent with the H1 data \cite{rhoLN_HERA}.
  
In Fig. \ref{fig:3} we present our predictions for the leading neutron spectra in  exclusive $\rho$ photoproduction considering different models for the 
pion flux. The corresponding $K$ values are also presented. We obtain a reasonable agreement with the experimental data, with $K$ values in the range 
$0.134 < K < 0.210$. It is important to emphasize that these values of $K$ are strongly correlated with our choice for $R_q$. For example, if instead of 
$R_q = 2/3$ we assume $R_q = 1/3$, the corresponding $K$ values should be multiplied by 4, since the exclusive cross sections depend  quadratically on the 
dipole scattering amplitude. If we assume {\it a priori} that the magnitude of the absorptive correction factor is of the 
order of $0.7$ for exclusive processes, as predicted in \cite{pirner} for semi - inclusive one, this implies a preference for the value $R_q = 1/3$. However, as 
the magnitude of these corrections for exclusive processes is still  an open question, as well as the value of $R_q$, 
we refrain from drawing strong conclusions. Therefore, in what follows we will only  present results assuming  $R_q = 2/3$, but the reader should keep in 
mind the quadratic  correlation between  $K$ and $R_q$, {  implying that the same fits could be obtained with much bigger values of $K$ and smaller values of $R_q$. 
With more data on different processes with leading neutron production, it may be possible to disentangle $K$ from $R_q$. 
 It is interesting to notice that leading neutron production is dominated by pion emission from the proton, 
i.e. $p \rightarrow  \pi  \, + \, n$. In all existing theoretical approaches this pion is soft and takes only a small fraction 
of  the incoming proton energy, leaving the neutron with most of it. This is the physical reason for the peak seen at $x_L \simeq 0.75$ 
in the $x_L$ spectrum of the leading neutron. 
}

In Fig. \ref{fig:4} we analyse the dependence of our predictions on the choice of the dipole scattering amplitude for two different models of the pion flux. 
As done before, we will constrain the value of $K$ by  adjusting the predictions of the different dipole models to the experimental value of the total cross 
section. We find that the different predictions for the $x_L$ spectra are very similar. However, the effective value of the absorptive correction $K$ depends 
on the model of the dipole scattering amplitude as expected, since they predict different values for the $B$ slope, which determines the normalization of the 
photon - pion cross section.  Figs. \ref{fig:3} and  \ref{fig:4} are in a sense complementary, since what changes in the former (the flux factor) is kept 
fixed in the latter (where the dipole model is changed) and vice-versa. In each curve the overall constant $ K R_q^2 $ is chosen so as to bring our calculations 
as close as possible to the experimental points. Comparing the curves we can conclude that the shape of the leading neutron spectrum is much more sensitive to the 
flux factor than to the dipole scattering amplitude   ${\mathcal N}$. The normalization of the spectrum is hence determined by $K$ and $R_q$, since  ${\mathcal N}$ 
is fixed from the analysis of other data. The values used for $K$ are significantly smaller than those found in theoretical estimates. Larger values of $K$ would be
more plausible implying a deviation from the valence quark scaling and the consequent change in the factor $2/3$. { A comparison between the curves in 
Fig. \ref{fig:4} favors the choice of the pion flux  $f_3$, which, unlike $f_2$, contains a t-dependent form factor. A similar preference  was found in 
Ref. \cite{china}, where a combined analysis of E866 and HERA data was performed. }

\begin{figure}[t]
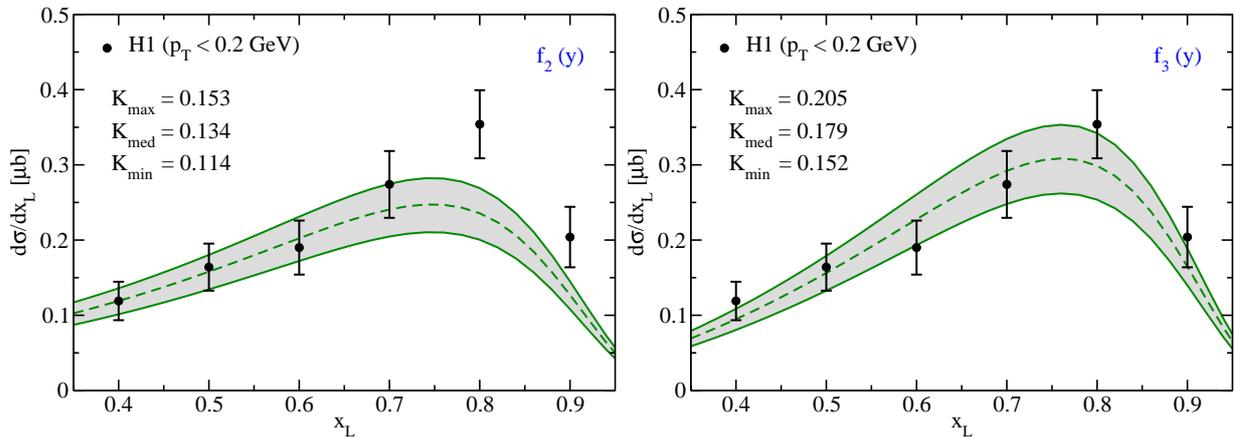

 \centering
 \includegraphics[scale=0.33]{pt0_2_bCGC_f2.eps}
 \includegraphics[scale=0.33]{pt0_2_bCGC_f3.eps}
 \caption{Leading neutron spectra in exclusive $\rho$ photoproduction obtained considering the possible range of values of the $K$ factor  and two models 
for the pion flux. Data from Ref. \cite{rhoLN_HERA}.}
 \label{fig:5}
\end{figure}

\begin{figure}[t]
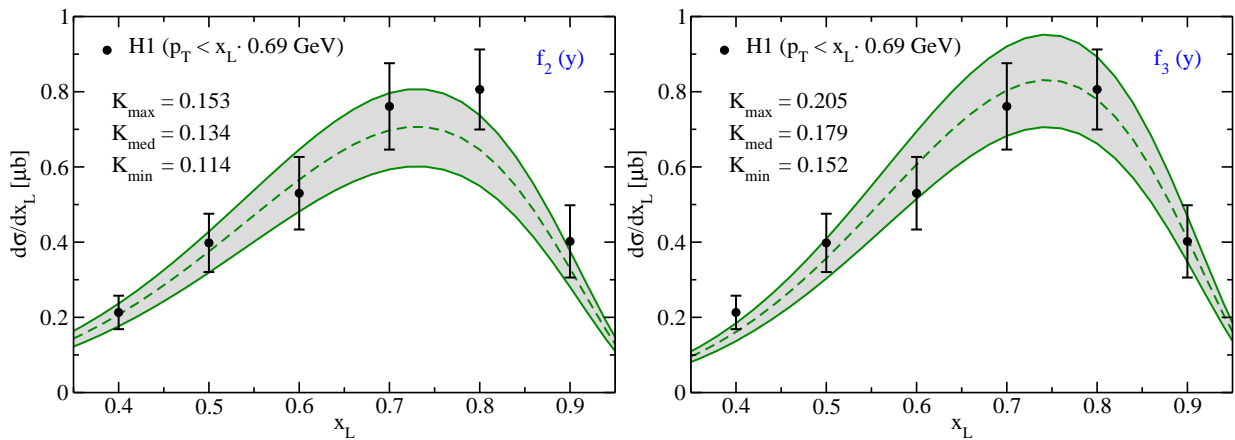

 \centering
 \includegraphics[scale=0.33]{pt0_69_bCGC_f2_K_0_2.eps}
 \includegraphics[scale=0.33]{pt0_69_bCGC_f3_K_0_2.eps}
 \caption{Leading neutron spectra in exclusive $\rho$ photoproduction obtained considering the possible range of values of the $K$ factor fixed using 
the other set of experimental data  and two models for the pion flux. H1 data \cite{rhoLN_HERA} obtained assuming that $p_T < 0.69 \cdot x_L$ GeV.}
 \label{fig:6}
\end{figure}

In what follows we will only consider the bCGC model, which successfully describes the HERA data on exclusive processes. In 
Fig. \ref{fig:5} we present our predictions taking into account the experimental uncertainty present in the H1 data for the total cross section 
\cite{rhoLN_HERA}, which in our analysis translates  into a range of possible values for the $K$ factor. These results indicate that the experimental data 
are better described using the pion flux $f_3 (y)$. As a cross check of our results, we can compare our predictions with other H1 data obtained assuming 
$p_T < 0.69 \cdot x_L$ GeV. Assuming the same range of values for $K$ obtained in Fig. \ref{fig:5} we can see in Fig. \ref{fig:6} that our  predictions 
describe  these data quite well.

\begin{figure}[t]
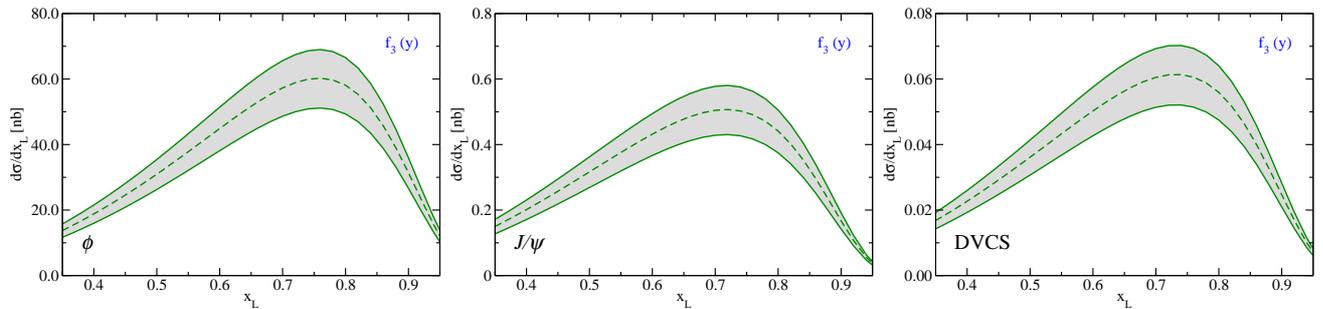

 \centering
 \includegraphics[scale=0.23]{bCGC_f3_phi.eps}
 \includegraphics[scale=0.23]{bCGC_f3_jpsi.eps}
 \includegraphics[scale=0.23]{bCGC_f3_dvcs.eps}
\caption{Predictions for the leading neutron spectra in  exclusive $\phi$, $J/\Psi$ and DVCS production in the HERA kinematical range: 
$W = 60$ GeV and $p_T < 0.2$ GeV. }
 \label{fig:7}
 \end{figure}

\begin{figure}[t]
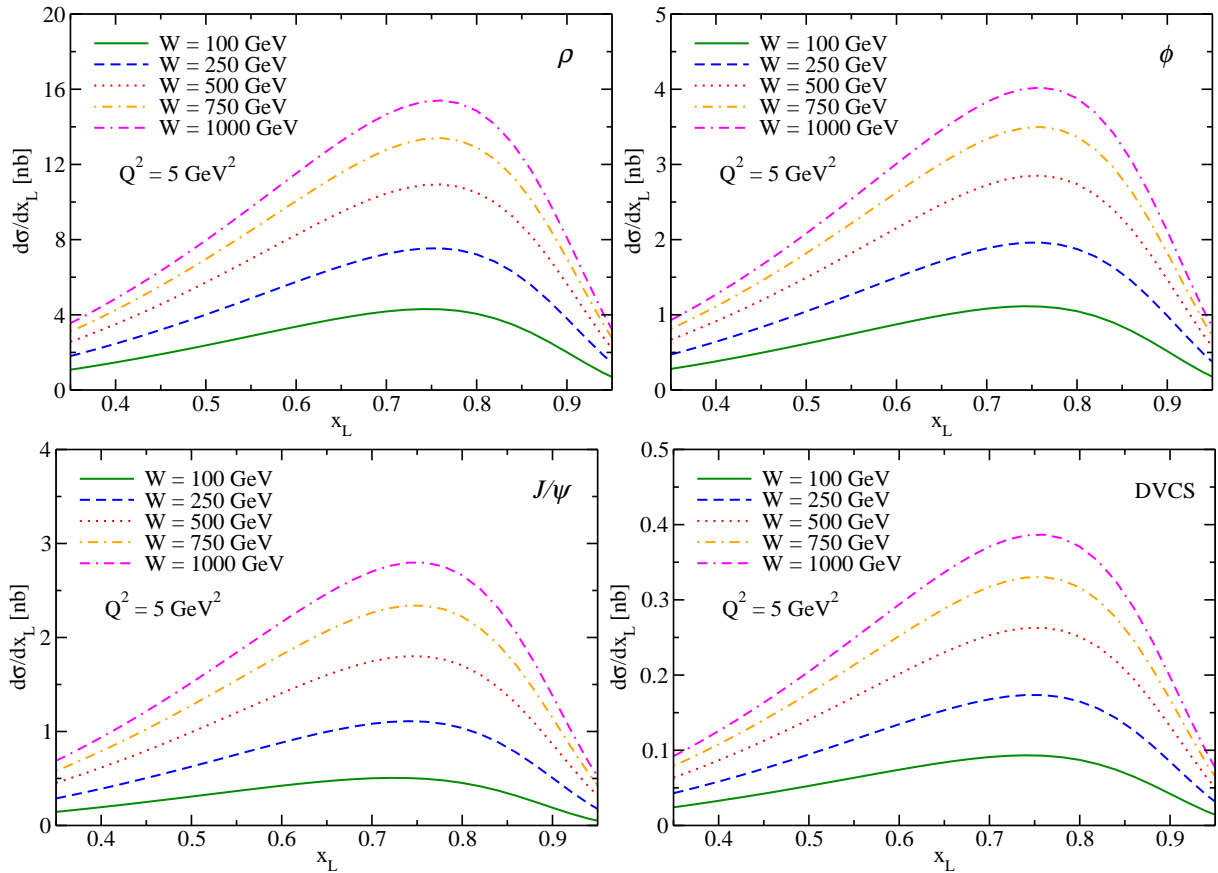

 \centering
 \includegraphics[scale=0.33]{rho_Q2_5_pt0_2_bCGC_f2.eps}
 \includegraphics[scale=0.33]{phi_Q2_5_pt0_2_bCGC_f2.eps}
 
 \includegraphics[scale=0.33]{jpsi_Q2_5_pt0_2_bCGC_f2.eps}
 \includegraphics[scale=0.33]{dvcs_Q2_5_pt0_2_bCGC_f2.eps}
 \caption{Energy dependence of the leading neutron spectra in exclusive $\rho$, $\phi$, $J/\Psi$ and DVCS production in the kinematical range 
of the future $ep$ colliders ($Q^2 = 5$ GeV$^2$). }
 \label{fig:8}
\end{figure}

\begin{figure}[t]
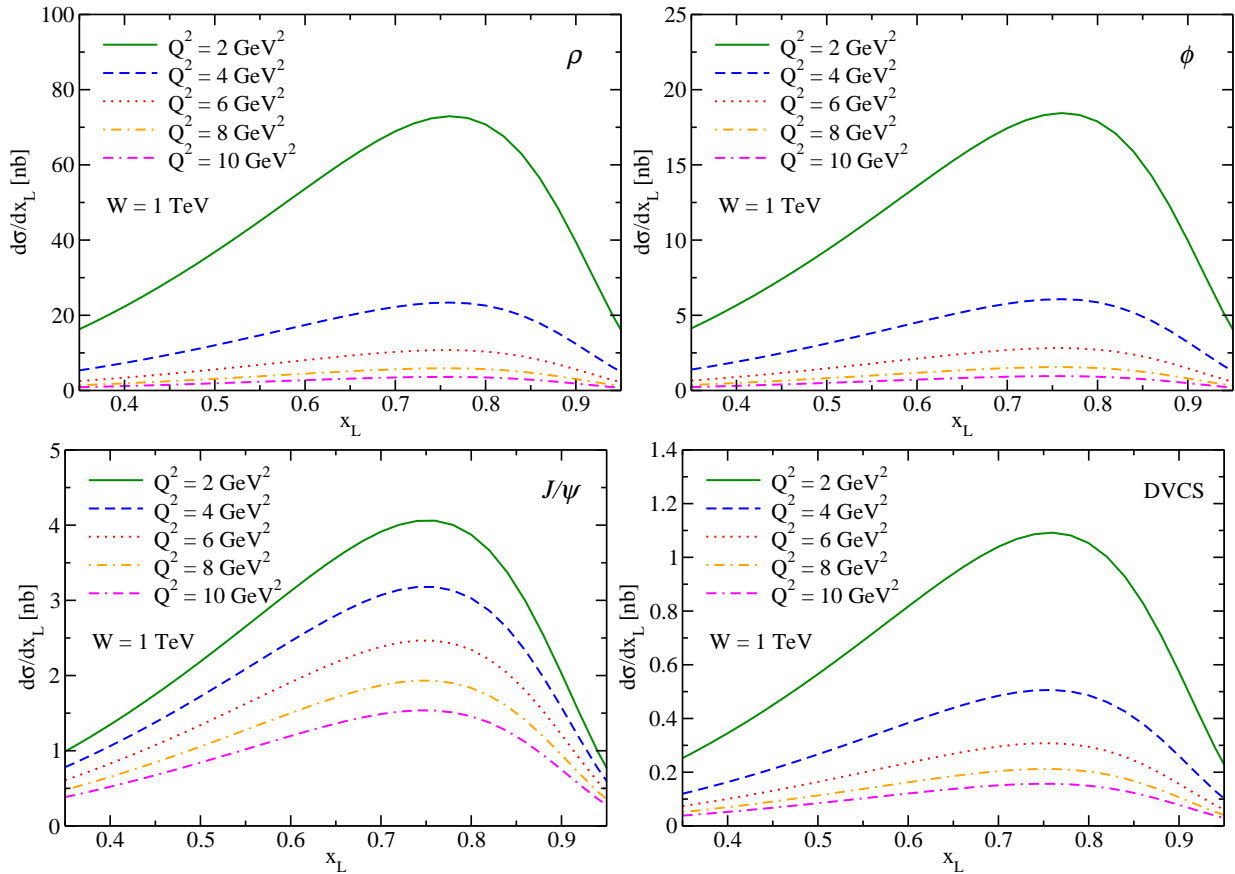

 \centering
 \includegraphics[scale=0.33]{rho_W_1TeV_pt0_2_bCGC_f2.eps}
 \includegraphics[scale=0.33]{phi_W_1TeV_pt0_2_bCGC_f2.eps}

 \includegraphics[scale=0.33]{jpsi_W_1TeV_pt0_2_bCGC_f2.eps}
 \includegraphics[scale=0.33]{dvcs_W_1TeV_pt0_2_bCGC_f2.eps}
 \caption{Dependence on the virtuality of the leading neutron spectra in exclusive $\rho$, $\phi$, $J/\Psi$ and DVCS production in the kinematical 
range of the future $ep$ colliders ($W = 1$ TeV).}
 \label{fig:9}
\end{figure}

Considering that the main inputs of our calculations have been fixed by the experimental data on  exclusive $\rho$ photoproduction we can extend our 
analysis to other exclusive final states. We will assume the Gauss-LC model for vector meson wave function, the bCGC dipole scattering amplitude, the 
$f_3$ model for the pion flux and the same $K$ values needed to describe the $\rho$ data.  Initially, let us consider the kinematical range probed by HERA. 
As in the $\rho$ case, we will assume $W = 60$ GeV and $p_T < 0.2$ GeV. However, for  $\phi$ and $J/\Psi$ production we assume $Q^2 = 0.04$ GeV$^2$, while 
for the DVCS we consider that $Q^2 = 10$ GeV$^2$. 
The corresponding predictions  for the leading neutron spectra in exclusive $\phi$ and $J/\Psi$  production as well as in   DVCS  are presented in Fig. \ref{fig:7}.
For the HERA kinematical range  we predict $\sigma (\gamma p \rightarrow  \phi \pi n) = 25.47 \pm 3.70$ nb,   
$\sigma (\gamma p \rightarrow  J/\Psi \pi n) = 0.22 \pm 0.03$ nb and $\sigma (\gamma^* p \rightarrow  \gamma \pi n) = 0.008 \pm 0.001$ nb, where the uncertainty 
is estimated taking into account the range of possible values of  $K$. Finally, let us present our predictions for the kinematical range which may  be probed 
in future $ep$ colliders assuming  $p_T < 0.2$ GeV. In Figs. \ref{fig:8} and \ref{fig:9}  we show our results for the energy  and photon virtualities, respectively.
As expected, the leading neutron spectra increases with the energy at fixed $Q^2$ and decreases with the virtuality at fixed $W$. In particular, for $W = 1$ 
TeV and $Q^2 = 5$ GeV$^2$ we predict $\sigma (\gamma^* p \rightarrow  \rho \pi n) = 6.55 \pm 0.95$ nb, 
$\sigma (\gamma^* p \rightarrow  \phi \pi n) = 1.71 \pm 0.25$ nb,   
$\sigma (\gamma^* p \rightarrow  J/\Psi \pi n) = 1.20 \pm 0.17$ nb and $\sigma (\gamma^* p \rightarrow  \gamma \pi n) = 0.16 \pm 0.02$ nb. We believe that 
for these values of total cross sections,  the experimental analysis of the exclusive processes associated with a leading neutron is feasible in  future $ep$ 
colliders. In particular, as the cross sections strongly increase when $Q^2 \rightarrow 0$, the analysis of the vector meson photoproduction in $ep$ collisions 
can be  useful to understand  leading neutron spectra, which are of crucial importance in particle production in cosmic ray physics. Another possibility is the 
study of this process in ultraperipheral hadronic collisions, with
 the leading neutron being a tag for  exclusive production.  In  principle these processes can be studied in the future at the LHC. Such proposition will be 
discussed in detail  in a forthcoming publication.

\section{Summary}

\label{conc}

One of the important goals in particle physics is to understand the production of leading particles, i.e. the production of   baryons 
which have large fractional longitudinal momentum ($x_L \geq 0.3$) and the same  valence quarks 
(or at least one of them) as the incoming particles. Recent measurements of   leading neutron spectra  in $ep$ collisions at HERA have shed a new light  
on this subject. However, the description of the semi - inclusive and exclusive leading neutron processes  remains without a satisfactory theoretical 
description. In a previous work \cite{nosLN}, we proposed to study  semi - inclusive leading neutron production using the color dipole 
formalism, which successfully describes both inclusive and diffractive HERA data, taking into account  the QCD dynamics and  its non - linear effects,  
which are expected to be present at high energies. Making use of  very simple assumptions about the relation between the dipole - pion and the dipole - proton 
scattering amplitudes and about the absorptive corrections, we demonstrated that the semi - inclusive data can be described by the dipole formalism and that  Feynman 
scaling is expected at high energies. In this paper we have extended our analysis to exclusive processes associated with a leading neutron. Considering the 
same assumptions used for the semi - inclusive case, we have analysed in detail the dependence of our predictions on the choices of the vector meson wave function, of 
the dipole model and and of the pion flux. We demonstrated that the HERA data on the exclusive $\rho$ photoproduction associated with a leading neutron can be 
quite well described by the color dipole formalism. Assuming the validity of this approach, we have presented for the first time predictions for the 
exclusive $\phi$, $J/\Psi$ and $\gamma$ production  in $ep$ collisions for the 
energies of HERA and future colliders. Our results indicate that the experimental analysis of these processes is feasible and that they can be used to understand 
this long standing problem in high energy physics.

{Finally, it is important to emphasize that the current sources of uncertainties  in the computation of leading neutron spectra are: 
i) the strength of the absorptive corrections represented by the factor $K$; 
ii) the validity of the additive quark model for the photon-pion cross section; 
iii) the strength of the contribution from direct fragmentation of the proton into neutrons; 
iv) the precise form of the pion flux; 
v) the precise form of the dipole cross section.
With sufficient  
experimental information, we can rule out  candidates of the pion flux and of the photon-pion cross section. 
We believe that it is possible to constrain  the unknown 
numbers and assumptions with the help of more 
experimental data on other processes  with tagged leading neutrons, such as those on $D^*$ production \cite{cheka-04}   
and those with dijet production \cite{cheka-10}. Work along this line is in progress.

}

\begin{acknowledgments}
This work was  partially financed by the Brazilian funding agencies CNPq, CAPES, FAPERGS and FAPESP.

\end{acknowledgments}

\hspace{1.0cm}

\end{document}